\documentclass[aps,pre,twocolumn,superscriptaddress,showpacs,showkeys]{revtex4}

\usepackage{natbib}
\usepackage{dcolumn}
\usepackage{amsmath}
\usepackage{graphicx}
\usepackage{rotating}
\usepackage{multirow}

\begin{document}

\title{Benchmark graphs for testing community detection algorithms}

\author{Andrea Lancichinetti}
\affiliation{Complex Systems Lagrange Laboratory (CNLL),
Institute for Scientific Interchange (ISI), Viale S. Severo 65, 10133, Torino, Italy}

\author{Santo Fortunato}
\affiliation{Complex Systems Lagrange Laboratory (CNLL),
Institute for Scientific Interchange (ISI), Viale S. Severo 65, 10133, Torino, Italy}

\author{Filippo Radicchi}
\affiliation{Complex Systems Lagrange Laboratory (CNLL),
Institute for Scientific Interchange (ISI), Viale S. Severo 65, 10133, Torino, Italy}

\date{\today} \widetext

\begin{abstract}

Community structure is one of the most important features of real networks and reveals
the internal organization of the nodes. Many algorithms have been
proposed but the crucial issue of testing, i.e. the 
question of how good an algorithm is, with respect to others, is still open. Standard tests include the analysis 
of simple artificial graphs with a built-in community structure, that the algorithm has to recover. However, the special 
graphs adopted in actual tests have a structure that does not reflect the real properties of nodes 
and communities found in real networks. Here we introduce a new class of benchmark graphs, that account for 
the heterogeneity in the distributions of node degrees and of community sizes. We use this new benchmark
to test two popular methods of community detection, modularity optimization and Potts model clustering. The results show that
the new benchmark poses a much more severe test to algorithms than standard benchmarks, revealing limits  
that may not be apparent at a first analysis.

\end{abstract}

\maketitle

\section{Introduction}
\label{sec1}

Many complex systems in nature, society and technology display a modular structure,
i.e. they appear as a combination of compartments that are fairly independent of each other.
In the graph representation of complex systems~\cite{Newman:2003,vitorep}, where the elementary units of a system are described as 
nodes and their mutual interactions as links, such modular structure is revealed by the existence of 
groups of nodes, called {\it communities} or {\it modules}, with many links connecting nodes of the same 
group and comparatively few links joining nodes of different groups~\cite{Girvan:2002,miareview}. 
Communities reveal a non-trivial internal organization of the network, and allow to infer special
relationships between the nodes, that may not be easily accessible from direct empirical tests.
Communities may be groups of related individuals 
in social networks~\cite{Girvan:2002, Lusseau:2005}, sets of Web pages dealing with the same topic~\cite{Flake:2002}, 
biochemical pathways in metabolic networks~\cite{Guimera:2005,palla}, etc. 

Detecting communities in networks is a big challenge. Many methods have been devised over the last few years,
within different scientific disciplines such as physics, biology, computer and social sciences. This race towards
the ideal method aims at two main goals, i.e. improving the accuracy in the determination of meaningful modules and 
reducing the computational complexity of the algorithm. The latter is a well defined objective: in many cases it 
is possible to compute analytically the complexity of an algorithm, in others one can derive it from simulations
of the algorithm on systems of different sizes. The main problem is then to estimate the accuracy of a method and to compare it with other methods.
This issue of testing is in our opinion as crucial as devising new powerful algorithms, but till now it has not received 
the attention it deserves. 

Testing an algorithm essentially means analyzing a network with a well defined community structure and 
recovering its communities. Ideally, one would like to have many instances of real networks whose modules
are precisely known, but this is unfortunately not the case. Therefore, the most extensive tests are performed
on computer generated networks, with a built-in community structure. The most famous benchmark for community detection
is a class of networks introduced by Girvan and Newman (GN)~\cite{Girvan:2002}. Each network has $128$ nodes, divided into four
groups with $32$ nodes each. The average degree of the network is $16$ and the nodes have approximately the same degree, as in a random
graph. At variance with a random graph, nodes tend to be connected preferentially to nodes of their group: a parameter $k_{out}$ indicates
what is the expected number of links joining each node to nodes of different groups (external degree). When $k_{out}<8$ each node
shares more links with the other nodes of its group than with the rest of the network. In this case, the four groups are well defined
communities and a good algorithm should be able to identify them.

This benchmark is regularly used to test algorithms. However, there are several caveats that one has to consider:
\begin{itemize}
\item{all nodes of the network have essentially the same degree;}
\item{the communities are all of the same size;}
\item{the network is small.}
\end{itemize}
The first two remarks indicate that the GN benchmark cannot be considered a proxy of 
a real network with community structure. 
\begin{figure*}[htb]
\includegraphics[width=\textwidth]{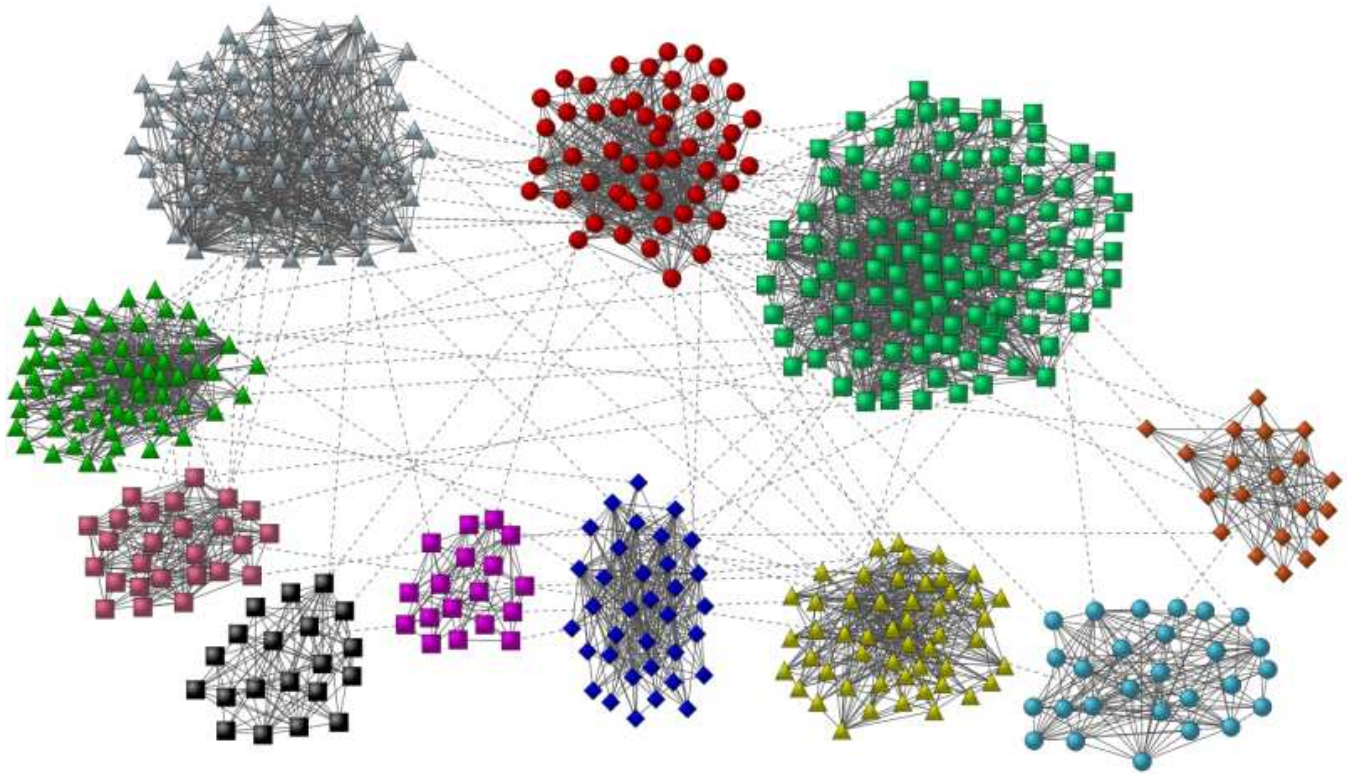}
\caption{\label{fig1} A realization of the new benchmark, with $500$ nodes.}
\end{figure*}
Real networks are characterized by heterogeneous
distributions of node degree, whose tails often decay as power laws. 
Such heterogeneity is responsible for a number of 
remarkable features of real networks, such as resilience 
to random failures/attacks~\cite{albert00}, and the absence of 
a threshold for percolation~\cite{cohen00} and epidemic spreading~\cite{pastor01}.
Therefore, a good benchmark should have a skewed degree distribution, like real networks.
Likewise, it is not correct to assume that all communities have the same
size: the distribution of community sizes of real networks is also broad, with a tail that can be fairly well approximated
by a power law~\cite{palla,guimera03,arenasrev,clausetfast}. A reliable benchmark should include communities of very different sizes.
A variant of the GN benchmark with communities of different size was introduced in~\cite{danon06}. 
Finally, the GN benchmark was a network of a reasonable size for most existing algorithms 
at the time when it was introduced. Nowadays, there are methods able to analyze graphs with 
millions of nodes~\cite{clausetfast,blondel08,lancichinetti08} and it is not appropriate to compare their performances on small graphs.
In general, an algorithm should be tested on benchmarks of variable size and average degree,
as these parameters may seriously affect the outcome of the method, and reveal its limits, as we shall see.

In this paper we propose a realistic benchmark for community detection, that accounts for the heterogeneity
of both degree and community size. Detecting communities on this class of graphs is a challenging task,
as shown by applying well known community detection algorithms.

\section{The benchmark}
\label{sec2}

We assume that both the degree and the community size distributions are power laws,
with exponents $\gamma$ and $\beta$, respectively. The number of nodes is $N$,
the average degree is $\langle k\rangle$. 

\begin{figure}[htb]
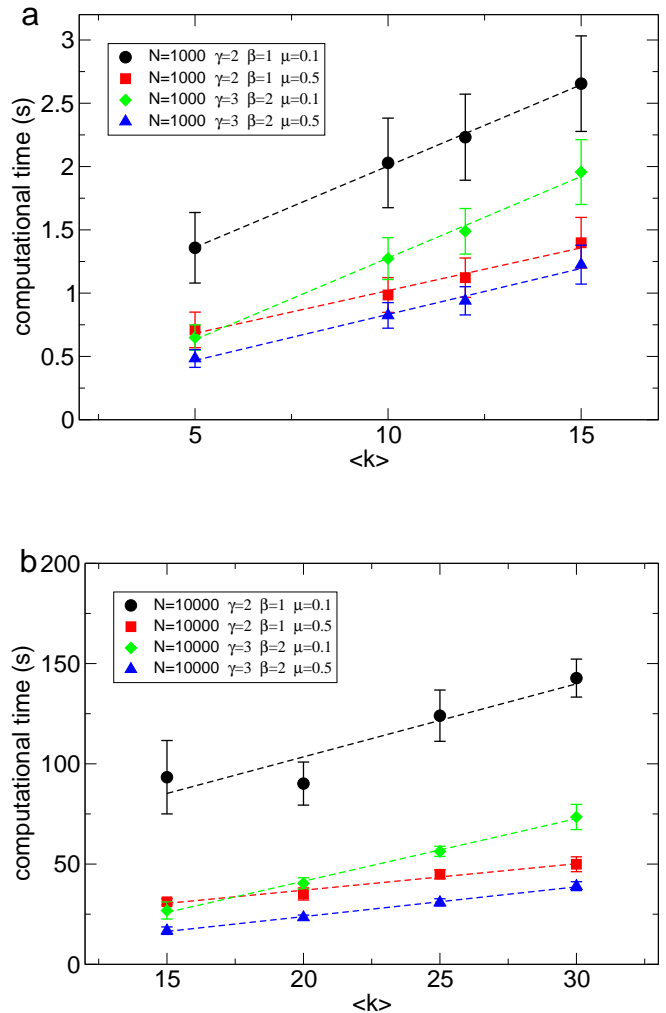

\begin{center}
\includegraphics[width=\columnwidth]{comptime.eps}
\vskip1cm
\includegraphics[width=\columnwidth]{comptime_a.eps}
\caption{\label{fig1b} Study of the complexity of our algorithm. The plots show the scaling of the computer time (in seconds) with
the average degree of the graph. The curves correspond to different choices for the exponents $\gamma$ and $\beta$ and the value of $\mu$. The two
panels reproduce graphs with $1000$ (a) and $10000$ nodes (b). The calculations were performed on Opteron processors.}
\end{center}
\end{figure}

In the GN benchmark a node may happen to have more links outside than inside its community even when
$k_{out}<8$, due to random fluctuations, which raises a conceptual problem concerning the natural classification of the node.
The construction of a realization of our benchmark proceeds through the following steps:
\begin{enumerate}
\item{Each node is given a degree taken from a power law distribution with exponent $\gamma$. The extremes of the distribution
$k_{min}$ and $k_{max}$ are chosen such that the average degree is $\langle k\rangle$. The configuration model~\cite{molloy}
is used to connect the nodes so to keep their degree sequence.}
\item{Each node shares a fraction $1-\mu$ of its links with the other nodes of its community and a fraction $\mu$ with the other nodes
of the network; $\mu$ is the {\it mixing parameter}.}
\item{The sizes of the communities are taken from a power law distribution with exponent $\beta$, such that the sum of all
sizes equals the number $N$ of nodes of the graph. The minimal and maximal community sizes $s_{min}$ and $s_{max}$ 
are chosen so to respect the constraints imposed by our definition of community: $s_{min} > k_{min}$ and $s_{max}>k_{max}$. This ensures that
a node of any degree can be included in at least a community.}
\item{At the beginning, all nodes are homeless, i.e. they are not assigned to any community.
In the first iteration, a node is assigned to a randomly chosen community; if the community size exceeds the internal degree
of the node (i.e. the number of its neighbors inside the community), the node enters the community, otherwise
it remains homeless. In successive iterations
we place a homeless node to a randomly chosen community: if the latter is complete, we kick out a randomly selected node
of the community, which becomes homeless. The procedure stops when there are no more homeless nodes.}
\item{To enforce the condition 
on the fraction of internal neighbors expressed by the mixing parameter $\mu$, several rewiring steps are performed, such that
the degrees of all nodes stay the same and only the split between internal and external degree is affected, when needed. In this way 
the ratio between external and internal degree of each node in its community can be set to the desired share $\mu$ with good approximation.}
\end{enumerate}
The prescription we have given leads to fast convergence.
In Fig.~\ref{fig1b} we show how the time to completion scales with the number of links of the graphs. The latter is expressed
by the average degree, as the number of nodes of the graphs is kept fixed. The curves clearly show a linear relation
between the computer time and the number of links of the graph. Therefore our procedure allows to build fairly large networks (up to $10^5-10^6$ nodes) 
in a reasonable time. 
Due to the strong constraints we impose to the system, in some instances convergence may not be reached.
However, this is very unlikely for the range of parameters we have used. For the exponents we have taken typical values of real networks:
$2\leq \gamma\leq 3$, $1\leq\beta\leq 2$. 

\begin{figure}[htb]
\begin{center}
\includegraphics[width=\columnwidth]{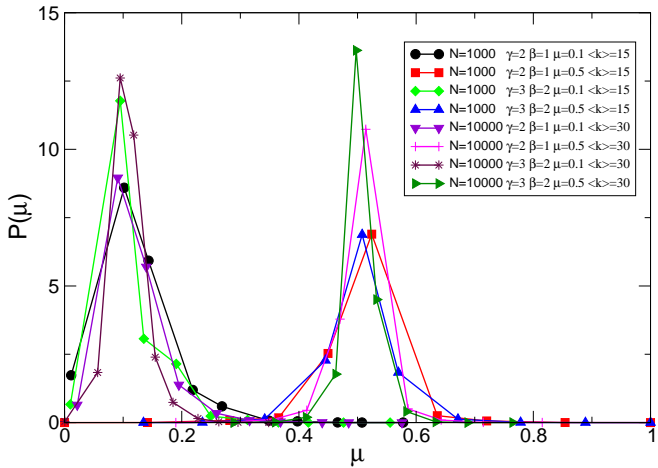}
\caption{\label{fig1c} Distribution of the $\mu$-values for benchmark graphs obtained with our algorithm for different choices of the exponents and system size. }
\end{center}
\end{figure}

Our algorithm tries to set the $\mu$-value of each node to the predefined input value, but of course this does not work in general,
especially for nodes
of small degree, where the possible values of $\mu$ are just a few and clearly separated. So, the distribution of $\mu$-values 
for a given benchmark graph cannot be a $\delta$-function, but it will have a bell-shaped curve, with a pronounced peak (Fig.~\ref{fig1c}).   

\section{Tests}
\label{sec3}

We have used our benchmark to test the performance of two methods to detect communities in networks,
i.e. modularity optimization~\cite{Newman:2004c,Duch:2005,Guimera:2005}, probably the most popular method of all, and the algorithm 
based on the Potts model introduced by Reichardt and Bornholdt~\cite{reichardt04}.

\begin{figure}[htb]
\begin{center}
\includegraphics[width=\columnwidth]{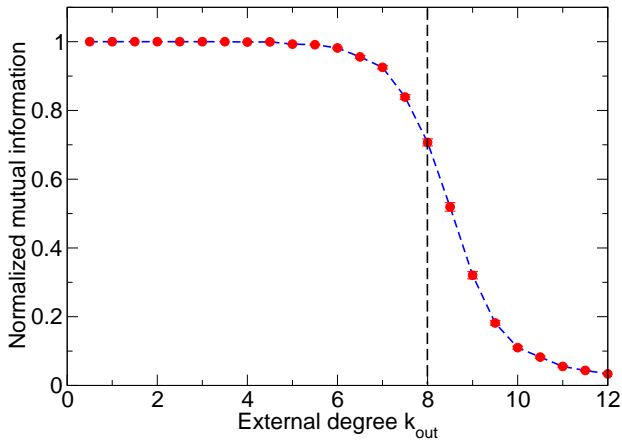}
\caption{\label{fig2} Test of modularity optimization on the benchmark of Girvan and Newman.}
\end{center}
\end{figure}

For modularity, the optimization was carried out 
through simulated annealing, as in~\cite{Guimera:2005}, which is not a fast technique but yields good estimates of modularity maxima.
In Fig.~\ref{fig2} we plot the performance of the method
as a function of the external degree of the nodes for the GN benchmark.
To compare the built-in modular structure with the one delivered by the algorithm
we adopt the {\it normalized mutual information}, a measure of 
similarity of partitions borrowed from information theory, which has proved to be 
reliable~\cite{Danon:2005}. As we can see from the figure, the natural partition is always found up until
$k_{out}=6$, then the method starts to fail, although it finds good partitions even when communities are fuzzy ($k_{out}\geq 8$).
Meanwhile, many algorithms are able to achieve comparable performances, so the benchmark can hardly discriminate between
different methods. As we can see from the figure, for $k_{out}< 8$ we are close to the top performance and there seems to be little room
for improvement. 

\begin{figure}[htb]
\begin{center}
\includegraphics[width=\columnwidth]{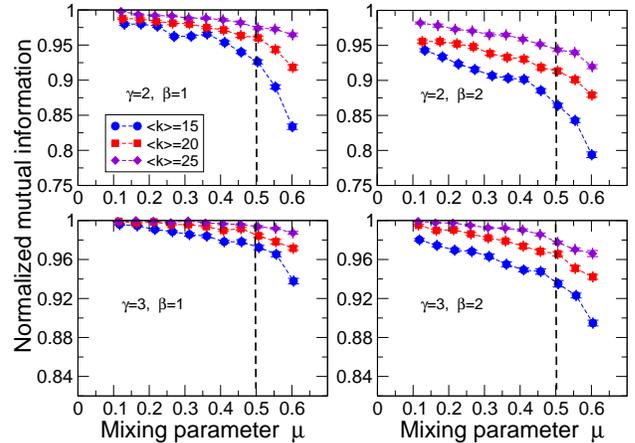}
\caption{\label{fig3} Test of modularity optimization on the new benchmark. The number of nodes $N=1000$. 
The results clearly depend on all parameters of the
benchmark, from the exponents $\gamma$ and $\beta$ to the average degree $\langle k \rangle$.
The threshold $\mu_c=0.5$ (dashed vertical line in the plots) 
marks the border beyond which communities are no longer defined in the strong sense, i.e. such that each 
node has more neighbors in its own community than in the others~\cite{radicchi}. Each point corresponds to an average over
$100$ graph realizations.}
\end{center}
\end{figure}

In Fig.~\ref{fig3} we show what happens if one optimizes modularity on the new benchmark, for $N=1000$. The four panels correspond to four pairs
for the exponents $(\gamma, \beta)=(2,1), (2,2), (3,1), (3,2)$. We have chosen combinations of the extremes of the exponents' ranges
in order to explore the widest spectrum of graph structures. For each pair of exponents, we have used three values for the average degree
$\langle k\rangle=15, 20, 25$. Each curve shows the variation of the normalized mutual information with the mixing parameter
$\mu$. 
\begin{figure}[htb]
\begin{center}
\includegraphics[width=\columnwidth]{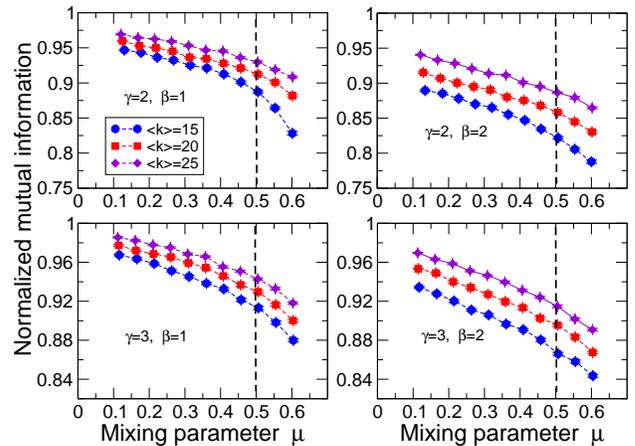}
\caption{\label{fig4} Test of modularity optimization on the new benchmark. The number of nodes is now $N=5000$, the other parameters
are the same as in Fig.~\ref{fig3}. Each point corresponds to an average over $25$ graph realizations.} 
\end{center}
\end{figure}
In general, from Fig.~\ref{fig3} we can infer that the method gives good results.
However, we find that it begins to fail even when communities are only loosely connected to each other
(small $\mu$). This is due to the fact that modularity optimization has an intrinsic resolution limit that makes small communities
hard to detect~\cite{FB}. Our benchmark is able to disclose this limit. We have explicitely verified that the modularity of the 
natural partition of the graph is lower than the maximum obtained from the optimization, and that the partition found by the algorithm has 
systematically a smaller number of clusters, due to the merge of small communities into larger groups.

We also see that the performance of the method is the better
the larger the average degree $\langle k\rangle$, whereas it gets worse when the communities are more similar to each other in size 
(larger $\beta$). 

\begin{figure}[htb]
\begin{center}
\includegraphics[width=\columnwidth]{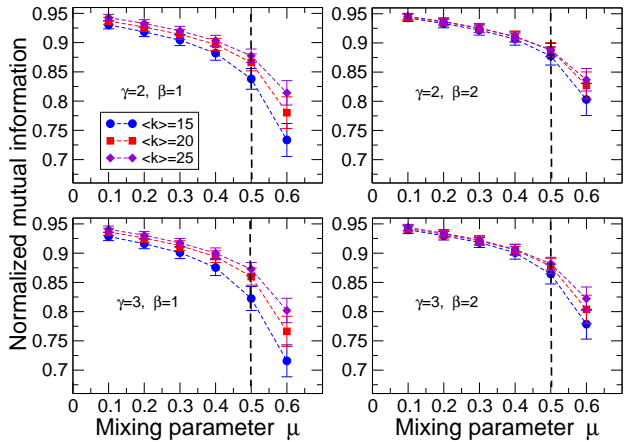}
\caption{\label{fig5} Test of Potts model clustering on the new benchmark. The number of nodes $N=1000$. 
The results clearly depend on all parameters of the
benchmark, from the exponents $\gamma$ and $\beta$ to the average degree $\langle k \rangle$. 
Each point corresponds to an average over $100$ graph realizations.}
\end{center}
\end{figure}
\begin{figure}[htb]
\begin{center}
\includegraphics[width=\columnwidth]{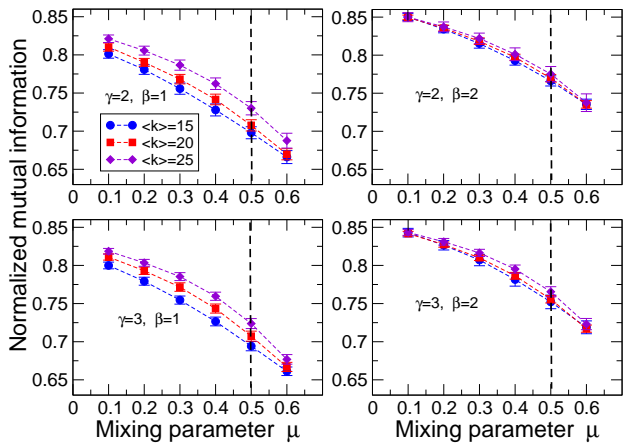}
\caption{\label{fig6} Test of Potts model clustering on the new benchmark. The number of nodes $N=5000$, 
the other parameters are the same as in Fig.~\ref{fig5}.  
Each point corresponds to an average over $10$ graph realizations.}
\end{center}
\end{figure}

To check how the performance is affected by the network size, we have tested the method on a set of larger graphs (Fig.~\ref{fig4}).
Now $N=5000$, whereas the other parameters are the same as before. Curves corresponding to the same parameters are similar,
but shifted towards the bottom for the larger systems. We conclude that the performance of the method worsens if the size of the
graph increases. If we consider that networks with $5000$ nodes are much smaller than many graphs one would like to analyze, 
modularity optimization may give inaccurate results in practical cases, something which could not be inferred from tests on existing benchmarks.

We have repeated the same analysis for the Potts model algorithm. We closely followed the implementation 
suggested by the authors of~\cite{reichardt04}: we set the number of spin states equal to the number of 
nodes of the network, the ferromagnetic coupling $J$ was set to $1$, whereas the antiferromagnetic coupling $\gamma$
equals the density of links of the network. The results are shown in Figs.~\ref{fig5} and ~\ref{fig6}. 
The performance of the method is fair, and it worsens for larger system sizes, like for modularity optimization, which proves superior.

\section{Summary}
\label{sec4}

We have introduced a new class of graphs to test algorithms identifying communities in networks. These new graphs extend the 
GN benchmark by introducing features of real networks, i.e. the heterogeneity in the distributions
of node degree and community size. We found that these elements pose a harder test to existing methods. We have tested modularity
optimization and a clustering technique based on the Potts model 
against the new benchmark. From the results the resolution limit of modularity emerges immediately. Furthermore, we have
seen that the size of the graph and the density of its links have a sizeable effect on the 
performance of the algorithm, so it is very important to study this dependence when testing a new algorithm.
The new benchmark is suitable for this type of analysis, as the graphs can be constructed very quickly, and one can span several
orders of magnitude in network size. A software package to generate the benchmark graphs can be downloaded
from {\it http://santo.fortunato.googlepages.com/benchmark.tgz}.

\end{document}